\documentstyle[prl,aps,amssymb,amsmath,epsf,psfrag,multicol]{revtex}

\begin{document}

\title{\large\bf Limitations on the Creation of Maximal Entanglement}
\author{Pieter Kok and Samuel L.\ Braunstein}
\address{Informatics, University of Wales, Bangor LL57 1UT, UK}

\maketitle

\begin{abstract}
 We study a limited set of optical circuits for creating near maximal 
 polarisation entanglement {\em without} the usual large vacuum contribution.
 The optical circuits we consider involve passive interferometers, feed-forward
 detection, down-converters and squeezers. For input vacuum fields we find
 that the creation of maximal entanglement using such circuits is impossible
 when conditioned on two detected auxiliary photons. So far, there have been
 no experiments with more auxiliary photons. Thus, based on the minimum
 complexity of the circuits required, if near maximal polarisation entanglement
 is possible it seems unlikely that it will be implemented experimentally with
 the current resources.
\end{abstract}

PACS number(s): 42.50.Dv

\begin{multicols}{2}
Entanglement is one of the key ingredients in quantum communication and 
information. For instance, quantum protocols such as dense coding, quantum 
error correction and quantum teleportation \cite{qic} rely on the non-classical
correlations provided by entanglement. Currently, substantial efforts are 
being made to use {\em optical} implementations for quantum communication.

The advantages of this are obvious: light travels at high speed and it weakly 
interacts with the environment. However, exactly this weak interaction poses 
serious drawbacks. The fact that photons do not interact with each other makes 
it hard to manipulate them. For example, it has recently been shown that it is 
impossible to perform so-called complete Bell measurements on two-mode 
polarisation states in linear quantum optics \cite{lutkenhaus1999,vaidman1999} 
(although theoretical schemes involving Kerr media \cite{kerr} and atomic 
coherence \cite{ac} have been reported). Furthermore, maximally 
polarisation-entangled two-photon states have not been produced. In this 
letter we investigate the possibility of creating such states with linear 
optics and a specific class of non-linear elements.

The maximally polarisation-entangled states which are most commonly considered 
are the Bell states:
\begin{eqnarray}\label{bellstates}
 |\Psi^{\pm}\rangle &=& \left( |\!\updownarrow\, ,\leftrightarrow\rangle \pm 
	|\!\leftrightarrow,\updownarrow\,\rangle \right) /\sqrt{2} \cr
 |\Phi^{\pm}\rangle &=& \left( |\!\updownarrow\, ,\updownarrow\,\rangle \pm 
	|\!\leftrightarrow,\leftrightarrow\rangle \right) /\sqrt{2} \; ,
\end{eqnarray}
where $|\!\!\updownarrow\,\rangle$ and $|\!\!\leftrightarrow\rangle$ 
denote single-photon states with orthogonal polarisations. In practice, these 
states have only been produced {\em randomly}, using for instance parametric 
down-conversion \cite{kwiat1995}. This process can yield a state 
\begin{equation}\label{pdc}
 |\psi\rangle \propto |0\rangle + \xi |\Psi^-\rangle + O(\xi^2)\; ,
\end{equation}
where $|0\rangle$ denotes the vacuum and $\xi\ll 1$. This means that the Bell 
state $|\Psi^-\rangle$ is only produced with a small probability of the order 
of $|\xi|^2$. Although $|\psi\rangle$ has a maximally entangled component, as 
a state it is very weakly entangled (this may be quantified by its partial Von 
Neumann entropy \cite{vonneuman}). Since we have no way of telling that an 
entangled photon-pair was produced without measuring (and hence destroying) 
the outgoing state, we call this randomly produced entanglement. Currently, in 
quantum optics we have access to this type of entanglement only.

By contrast, we would like to be able to tell that we in fact produced a
maximally entangled state before it is used. That is, we wish to have a source 
which gives a macroscopic indication that a maximally polarisation-entangled 
state has been produced. Such a source is said to create {\em event-ready} 
entanglement. The vacuum contribution in Eq.\ (\ref{pdc}) can be eliminated
by means of a polarisation independent QND measurement. However, this would 
involve higher-order non-linearities (like the Kerr effect) which, in 
practice, are very noisy (especially when they are required to operate at the 
single-photon level). In general, the creation of event-ready entanglement 
can be quantified by a certain probability of `happening'. When this 
probability is equal to one, we have a {\em deterministic} source of 
event-ready entanglement. 

Random entanglement has been used to demonstrate, for example, non-local
features of quantum teleportation and entanglement swapping 
\cite{bouwmeester1997,pan1998,kok}. One 
might therefore suppose that in practice we don't really need event-ready 
entanglement. However, on a theoretical level Bell states appear as primitive 
notions. This means that protocols like entanglement purification and error 
correction \cite{bennett1996a,bennett1996} have been designed for maximally 
entangled states, rather than for random entanglement. For quantum 
communication to become a mature technology, one most certainly needs the 
ability to perform entanglement purification and error correction. It is not 
at all clear how these protocols can be convincingly implemented with random 
entanglement. One approach would be to try and investigate such protocols.
However, that is not our aim here.

In this letter we give limitations to the creation of near maximal entanglement
with linear optics and some non-linear optical components (such as 
down-converters and squeezers). First we present the tools with which we will
attempt to produce event-ready entanglement. Then we derive a general condition
for an optical setup, which should be satisfied in order to yield event-ready 
entanglement. We subsequently examine this condition in the context of several
types of photon-sources.

Given a pair of photons in one maximally polarisation-entangled state, we can 
obtain any other such state by a combination of a polarisation rotation and a 
polarisation dependent phase shift. When we study the creation of maximal 
polarisation-entanglement we shall therefore restrict ourselves to the 
$|\Psi^-\rangle$ Bell state without loss of generality. 

\begin{figure}[t]
\label{fig1}
\begin{center}
  \begin{psfrags}
     \psfrag{a}{a)}
     \psfrag{b}{b)}
     \psfrag{psi}{$|\Psi^-\rangle$}
     \psfrag{u0}{$U^{(0)}$}
     \psfrag{u1}{$U^{(1)}$}
     \psfrag{un}{$U^{(n)}$}
     \psfrag{u}{$U$}
     \psfrag{in}{$|\psi_{\rm in}\rangle$}
     \psfrag{to}{$\Longrightarrow$}
     \psfrag{d}{\ldots}
     \psfrag{b5}{$a_5$}
     \psfrag{bn}{$a_N$}
     \epsfxsize=8in
     \epsfbox[-10 20 1000 270]{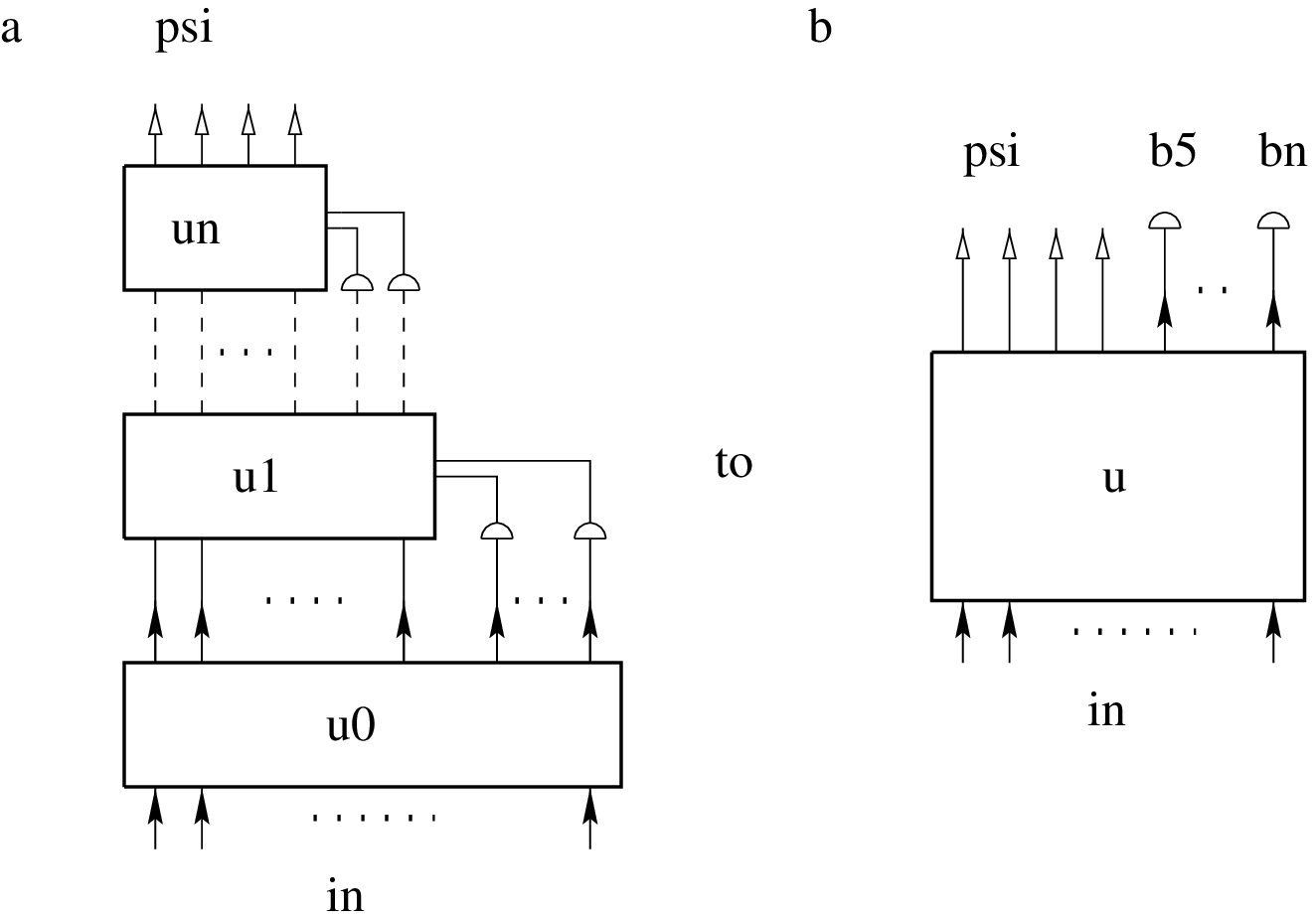}
  \end{psfrags}
  \end{center}
  {\small Fig.\ 1: If an optical circuit with feed-forward detection (a) 
	produces a specific state, the same output can be obtained by an 
	optical circuit where detection of the {\em auxiliary} modes takes 
	place at the end (b). The efficiency of the latter, however, will 
	generally be smaller.}
\end{figure}

In order to make $|\Psi^-\rangle$, we will assume that we have several 
resources at our disposal. {\em In this letter, the class of reasonable
elements will consist of beam-splitters, phase-shifters, photo-detectors and 
non-linear components such as down-converters, squeezers, etc.} These elements 
are then arranged to give a specific optical circuit (see Fig.\ 1a). Part of 
this setup might be so-called {\em feed-forward} detection. In this scheme the 
outcome of the detection of a number of modes dynamically chooses the internal 
configuration of the subsequent optical circuit based on the interim detection 
results (see also Ref.\ \cite{lutkenhaus1999}). Conditioned on these 
detections we want to obtain a freely propagating $|\Psi^-\rangle$ Bell state 
in the remaining undetected modes. 

We now introduce two simplifications for such an optical circuit. First, we 
will show that we can discard feed-forward detection. Secondly, we will see 
that we only have to consider the detection of modes with at most one photon.

{\bf Theorem 1:} In order to show that it is {\em possible} to produce a 
specific outgoing state, any optical circuit with feed-forward detection can 
be replaced by a {\em fixed} optical circuit where detection only takes place 
at the end.

{\bf Proof:} Suppose a feed-forward optical circuit (like the one depicted in 
Fig.\ 1a) giving $|\Psi^-\rangle$ exists. That means that the circuit creates
$|\Psi^-\rangle$ conditioned on one of potentially many patterns of detector
responses. It is sufficient to consider a single successful pattern. We can 
then take every interferometer to be fixed and postpone all detections of the 
auxiliary modes to the very end (Fig.\ 1b). Note that this procedure selects 
generally only {\em one} setup in which entanglement is produced, whereas a 
feed-forward optical circuit potentially allows more setups. It therefore 
might reduce the efficiency of the process. However, since we are only 
interested in the {\em possibility} of creating $|\Psi^-\rangle$, the 
efficiency is irrelevant.\hfill $\square$

{\bf Theorem 2:} Suppose an optical circuit produces a specific outgoing state 
conditioned on $n_1$ detected photons in mode 1, $n_2$ detected photons in 
mode 2, etc.\ (with $n_i=0,1,2,\ldots$). The same output can be obtained by 
a circuit where in every detected mode {\em at most} one photon is found.

{\bf Proof:} If there are more photons in a mode, we can replace the 
corresponding detector by a so-called detector {\em cascade} \cite{kok2}. This 
device splits the mode into many modes which are all detected. For a 
sufficiently large cascade there is always a non-vanishing probability to have 
at most one photon in each outgoing mode. In that case, the same state is 
created while at most one photon enters each detector. Note that this again 
yields a lower efficiency. \hfill $\square$

Applying these results to the creation of $|\Psi^-\rangle$, it is sufficient 
to consider a single {\em fixed} interferometer acting on an incoming state, 
followed at the end by detection of the so-called {\em auxiliary} modes. 
$|\Psi^-\rangle$ is signalled by at least one fixed detection pattern with at 
most one photon in each detector.

How do we proceed in trying to make the $|\Psi^-\rangle$ Bell state? Let the 
time independent interaction Hamiltonian ${\mathcal{H}}_I$ incorporate both the
interferometer $U$ and the creation of $|\psi_{\rm in}\rangle$ (see Fig.\ 1b). 
The outgoing state {\em prior to the detection} can be formally written as
\begin{equation}\label{evolution}
 |\psi_{\rm out}\rangle = U |\psi_{\rm in}\rangle \equiv \exp\left( -it 
 {\mathcal{H}}_I/\hbar \right) |0\rangle\; ,
\end{equation}
with $|0\rangle$ the vacuum. This defines an effective Hamiltonian 
${\mathcal{H}}_I$ which is generally not unique.

At this point we find it useful to change our description. Since the creation 
and annihilation operators satisfy the same commutation relations as c-numbers
and their derivatives, we can make the substitution $a^{\dagger}_i\rightarrow
\alpha_i$ and $a_i\rightarrow\partial_i$, where $\partial_i\equiv\partial/
\partial\alpha_i$. Furthermore, we define $\vec\alpha = (\alpha_1,\ldots,
\alpha_N)$. Quantum states are then represented by functions of c-numbers and 
their derivatives. This is called the Bargmann representation \cite{bargmann}.

Furthermore, suppose we can normal order the operator $\exp(-it{\mathcal{H}}_I 
/\hbar)$ in Eq.\ (\ref{evolution}). This would leave us with a function of 
only the creation operators, acting on the vacuum. In the Bargmann 
representation we then obtain a function of complex numbers without their 
derivatives. In particular, when we have an optical circuit consisting of $N$ 
distinct modes (for notational convenience we treat distinct polarisations 
like, for instance, $x$ and $y$ as separate modes), we obtain the function
$\psi_{\rm out} (\vec{\alpha})$ after the unitary evolution $U$ and normal 
ordering. The normal ordering of the evolution operator in conjunction with 
the vacuum input state is crucial, since it allows us to simplify the problem 
significantly.

We now treat the (ideal) detection of the auxiliary modes in the Bargmann 
representation. Suppose the outgoing state after the detection of $M$ 
photons emerges in modes $a_1$, $a_2$, $a_3$ and $a_4$. After a suitable 
reordering of the detected modes the state which is responsible for the 
detector coincidence indicating success can be written as $|1_5,\ldots\!,
1_{M+4}, 0_{M+5},\ldots\rangle$ (possibly on a countably infinite number of 
modes). We then have the post-selected state $|\psi_{\rm post}\rangle$
\begin{eqnarray}\label{detcs}
 |\psi_{\rm post}\rangle_{1..4} &\propto& \langle 1_5,\ldots\!,1_{M+4}, 
 0_{M+5}, \ldots| \psi_{\rm out}\rangle\cr 
 &=& \langle 0|\, a_5 \cdots a_{M+4} |\psi_{\rm out}\rangle\; .
\end{eqnarray}
In the Bargmann representation the right-hand side of Eq.\ (\ref{detcs}) is
\begin{equation}
 \left. \partial_5 \cdots \partial_{M+4}\; \psi_{\rm out} (\vec\alpha) 
 \right|_{\vec\alpha' = 0} \; ,
\end{equation}
where we write $\vec\alpha' = (\alpha_5,\ldots\!,\alpha_{M+4},\ldots)$.

Writing out the entanglement explicitly in the four modes (treating the 
polarisation implicitly), we arrive at the following condition for the 
creation of two photons in the antisymmetric Bell state:
\begin{equation}\label{con}
 \left. \partial_5 \cdots \partial_{M+4}\; \psi_{\rm out} (\vec\alpha) 
 \right|_{\vec\alpha'=0} \propto \alpha_1 \alpha_2 - \alpha_3 \alpha_4 + 
 O(\xi)\; .
\end{equation}
The term $O(\xi)$ will allow for a small pollution ($\xi\ll 1$) in the 
outgoing state. We will show that for certain special classes of interaction 
Hamiltonians this condition is very hard (if not impossible) to satisfy. This 
renders the experimental realisation of two maximally polarisation entangled 
photons at least highly impractical.

We are now ready to shape $\psi_{\rm out}$ in more detail. In this letter we
consider two distinct classes of interaction Hamiltonians. 

First, suppose ${\mathcal{H}}_I$ is linear in the creation operators. This 
means that the optical circuit consists of coherent inputs, linear operations 
and no squeezing. The state prior to the detection can be written as 
$\exp(\sum_i d_i \alpha_i)$. We immediately see that the detection in 
condition (\ref{con}) only yields constant factors. This can never give us 
the $|\Psi^-\rangle$ Bell state. 

By contrast, we consider optical circuits including mode-mixing, squeezers and 
down-converters. The corresponding interaction Hamiltonians ${\mathcal{H}}_I$ 
are {\em quadratic} in the creation operators. There are no linear terms, so
there are no coherent displacements. More formally
\begin{equation}\label{source}
 {\mathcal{H}}_I = \sum_{i,j=1}^N a^{\dagger}_i A^{(1)}_{ij} a^{\dagger}_j + 
 \sum_{i,j=1}^N a^{\dagger}_i A^{(2)}_{ij} a_j + 
 {\rm H.c.}\; .
\end{equation}
With $A^{(1)}$ and $A^{(2)}$ complex matrices.
According to Braunstein \cite{blubber}, such an active interferometer is 
equivalent to a passive interferometer $V$, followed by a set of single-mode 
squeezers and another passive interferometer $U'$. We can view the photon 
source described by Eq.\ (\ref{source}) as an active bilinear component of an 
interferometer. For vacuum input and after normal ordering \cite{truax}, the 
optical setup then gives rise to
\begin{equation}\label{bilinear}
 \psi_{\rm out} = \exp\left[(\vec\alpha,B\vec\alpha)\right]\; ,
\end{equation} 
with $(\vec\alpha,B\vec\alpha)=\sum_{ij}^N \alpha_i B_{ij} \alpha_j$.
Such an optical setup would correspond to a collection of single-mode squeezers
acting on the vacuum, followed by a passive optical interferometer $U'$. Here, 
$B$ is a complex symmetric matrix determined by the interaction Hamiltonian 
${\mathcal{H}}_I$ and the interferometer $U'$. We take $B$ to be proportional 
to a common coupling constant $\xi$. The outgoing auxiliary modes $a_5$ to 
$a_N$ are detected (see Fig.\ 2). We will now investigate whether we can 
produce $|\Psi^-\rangle$ conditioned on a given number of detected photons.

\begin{figure}[t]
\label{fig2}
\begin{center}
  \begin{psfrags}
     \psfrag{5}{$a_5$}
     \psfrag{N}{$a_N$}
     \psfrag{u}{$U'$}
     \psfrag{s}{s}
     \psfrag{in}{$|0\rangle$}
     \psfrag{out}{$|\Psi^-\rangle$}
     \psfrag{un}{$\underbrace{\phantom{xxxxxxxxxxxxxx.}}$}
     \epsfxsize=8in
     \epsfbox[-108 20 700 150]{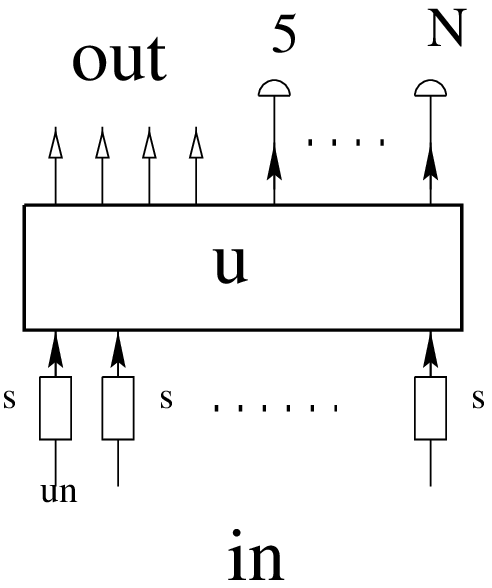}
  \end{psfrags}
  \end{center}
  {\small Fig.\ 2: The unitary interferometer $U'$ with conditional 
	photo-detection and single-mode squeezers which should transform 
	$|0\rangle$ into $|\Psi^-\rangle$.}
\end{figure}

In the case of a bilinear interaction Hamiltonian, photons are always created 
in pairs. In addition, we seek to create {\em two} maximally entangled photons.
An odd number of detected photons can never give $|\Psi^-\rangle$ and the 
number of detected photons should therefore be even. The lowest even number is
zero. In this case no photons are detected and $\psi_{\rm out}$ in Eq.\ 
(\ref{bilinear}) is proportional to $1+O(\xi)$, which corresponds to the 
vacuum state.

The next case involves two detected photons. To have entanglement in modes 
$\alpha_1$ to $\alpha_4$ after detecting {\em two} photons requires
\begin{equation}\label{2con}
 \left. \partial_5 \partial_6 \; e^{(\vec\alpha, B \vec\alpha)}
 \right|_{\vec\alpha'=0} \propto \alpha_1\alpha_2-\alpha_3\alpha_4 + O(\xi)\; .
\end{equation}

The left-hand side of Eq.\ (\ref{2con}) is equal to 
\begin{equation}
 \Bigl( B_{56} + \sum_{i,j=1}^4 \alpha_i B_{i5} B_{j6} \alpha_j \Bigr) \left.
 e^{(\vec\alpha,B \vec\alpha)} \right|_{\vec\alpha'=0} \; .
\end{equation}
To satisfy Eq.\ (\ref{2con}), the vacuum contribution $B_{56}$ should be 
negligible. We now ask whether the second term can give us entanglement.

The right hand side of Eq.\ (\ref{2con}) can be rewritten according to
$\alpha_1\alpha_2 - \alpha_3 \alpha_4 = \sum_{i,j=1}^4 \alpha_i E_{ij} 
\alpha_j$, where $E_{ij}$ are the elements of a symmetric matrix $E$. It is 
easily seen that $\det E = 1$. 

Let $M_{ij} = B_{i5}B_{j6}$. Since only the symmetric part of $M$ contributes,
we construct $\widetilde{M}_{ij} = (M_{ij} + M_{ji})/2$. The condition for
two detected photons now yields
\begin{equation}
 \sum_{i,j=1}^4 \alpha_i \widetilde{M}_{ij} \alpha_j = \sum_{i,j=1}^4
 \alpha_i E_{ij} \alpha_j + O(\xi)\; ,
\end{equation}
If this equality is to hold, we would need $\det E = \det\widetilde{M} + 
O(\xi) = 1$. However, it can be shown that $\det\widetilde M = 0$. $\widetilde 
M$ can therefore never have the same form as $E$ for small $\xi$, so it is not 
possible to create maximal polarisation entanglement conditioned upon two 
detected photons.

The last case we consider here involves four detected photons.
When we define $X_i = \sum_{j} B_{ij} \alpha_j$, the left-hand side of Eq.\ 
(\ref{con}) for four detected photons gives
\begin{eqnarray}\nonumber
 \left( \right. && B_{56} B_{78} + B_{57} B_{68} + B_{58} B_{67} + 
	B_{56} X_7 X_8 + \cr
 && \left. B_{57} X_6 X_8 + B_{58} X_6 X_7 + B_{67} X_5 X_8 + B_{68} X_5 X_7 + 
	\right. \cr
 && \left. B_{78} X_5 X_6 + X_5 X_6 X_7 X_8 \right)\left. 
	e^{(\vec\alpha,B\vec\alpha)}\right|_{\vec\alpha'=0}
	\; .
\end{eqnarray}
We have not been able either to prove or disprove that $|\Psi^-\rangle$ can be 
made this way. The number of terms which contribute to the bilinear part in 
$\alpha$ rapidly increases for more detected photons. 

We have proved that multi-mode squeezed vacuum conditioned on two detected 
photons cannot give maximal entanglement. We conjecture that this is true for
any number of detected photons. However, suppose we {\em could} create
maximal entanglement conditioned upon four detected photons, how efficient 
would this process be? For four detected photons yielding $|\Psi^-\rangle$ we 
need at least three photon-pairs. These are created with a probability of the 
order of $|\xi|^6$. Currently, $|\xi|^2$, the probability per mode, has a 
value of $10^{-4}$ \cite{weinfurter}. For experiments operating at a 
repetition rate of 100 MHz using ideal detectors, this will amount to 
approximately one maximally entangled pair every few hours. For realistic 
detectors this is much less. 

So far, there have been no experiments which exceeded the detection of more 
than two {\em auxiliary} photons (not including the actual detection of the 
maximally entangled state). This, and the estimation of the above efficiency 
appears to place strong practical limitations on the creation of maximal 
entanglement. 

In this letter we have demonstrated strong limitations on the possibility of
creating maximal entanglement with quantum optics. To this end, we introduced
two simplifications to our hypothetical optical circuit: we replaced 
feed-forward detection by a fixed set of detectors at the end, and secondly, 
every detector needs to detect at most one photon. Conditioned on two detected 
photons, multi-mode squeezed vacuum fails to create maximal entanglement. This 
leads to our conjecture that maximal entanglement is impossible using only 
these sources and linear interferometry. There is a number of open questions. 
First of all, is our conjecture true? And secondly, what happens when we have 
a combination of squeezing and coherent displacements? In that case the 
approach taken here fails due to the more complex normal ordering of the 
interaction Hamiltonian.

Entanglement is a fascinating and important phenomenon in physics. It not 
only provides us with insights in the mysterious world of quantum mechanics,
but it also appears as a fundamental resource in quantum information and
communication theory. However, {\em maximal polarisation} entanglement has 
never been produced in the laboratory. We have shown here the this type of
entanglement proves to be highly elusive. 

SLB would like to thank Klaus M\o lmer for valuable discussions. This research 
is funded in part by EPSRC grant GR/L91344.

\end{multicols}

\end{document}